\documentclass[twocolumn,english,prl,floatfix,showpacs,amsmath,amssymb,superscriptaddress]{revtex4}
\usepackage{mathptmx}

\usepackage[T1]{fontenc}
\usepackage[latin9]{inputenc}
\usepackage{bm}
\usepackage{amsmath}
\usepackage{amssymb}
\usepackage{graphicx}

\makeatletter
\@ifundefined{textcolor}{}
{%
 \definecolor{BLACK}{gray}{0}
 \definecolor{WHITE}{gray}{1}
 \definecolor{RED}{rgb}{1,0,0}
 \definecolor{GREEN}{rgb}{0,1,0}
 \definecolor{BLUE}{rgb}{0,0,1}
 \definecolor{CYAN}{cmyk}{1,0,0,0}
 \definecolor{MAGENTA}{cmyk}{0,1,0,0}
 \definecolor{YELLOW}{cmyk}{0,0,1,0}
 }

\usepackage{bm}% bold math

\newcommand{\bP}{\mathbf{P}}
\newcommand{\bnot}{\mathbf{0}}
\newcommand{\be}{\begin{equation}}
\newcommand{\ee}{\end{equation}}
\newcommand{\bea}{\begin{eqnarray}}
\newcommand{\eea}{\end{eqnarray}}
\newcommand{\ba}{\begin{eqnarray*}}
\newcommand{\ea}{\end{eqnarray*}}
\newcommand{\dagga}{{\phantom{\dagger}}}
\newcommand{\bR}{\mathbf{R}}

\newcommand{\bq}{\mathbf{q}}

\newcommand{\br}{\mathbf{r}}

\usepackage{babel}

\makeatother

\usepackage{babel}
\begin{document}

\title{Light-Cone Effect and Supersonic Correlations in One and Two-Dimensional
Bosonic Superfluids}

\author{Giuseppe Carleo}

\affiliation{Laboratoire Charles Fabry, Institut d'Optique, CNRS, Univ. Paris
Sud 11, 2 avenue Augustin Fresnel, F-91127 Palaiseau cedex, France }

\author{Federico Becca}

\affiliation{Democritos Simulation Center CNR-IOM Istituto Officina dei Materiali
and International School for Advanced Studies (SISSA), Via Bonomea
265, 34136 Trieste, Italy }

\author{Laurent Sanchez-Palencia}

\affiliation{Laboratoire Charles Fabry, Institut d'Optique, CNRS, Univ. Paris
Sud 11, 2 avenue Augustin Fresnel, F-91127 Palaiseau cedex, France }

\author{Sandro Sorella}

\affiliation{Democritos Simulation Center CNR-IOM Istituto Officina dei Materiali
and International School for Advanced Studies (SISSA), Via Bonomea
265, 34136 Trieste, Italy }

\author{Michele Fabrizio}

\affiliation{Democritos Simulation Center CNR-IOM Istituto Officina dei Materiali
and International School for Advanced Studies (SISSA), Via Bonomea
265, 34136 Trieste, Italy }

\date{\today}
\begin{abstract}
We study the spreading of density-density correlations in Bose-Hubbard
models after a quench of the interaction strength, using time-dependent
variational Monte Carlo simulations. It gives access to unprecedented
long propagation times and to dimensions higher than one. In both
one and two dimensions, we find ballistic light-cone spreading of
correlations and extract accurate values of the light-cone velocity
in the superfluid regime. We show that the spreading of correlations
is generally supersonic, with a light-cone propagating faster than
sound modes but slower than the maximum group velocity of density
excitations, except at the Mott transition, where all the characteristic
velocities are equal. Further, we show that in two dimensions the
correlation spreading is highly anisotropic and presents nontrivial
interference effects. 
\end{abstract}


\maketitle
\textit{Introduction.---} In 1972 Lieb and Robinson demonstrated that
an effective light cone emerges in non-relativistic quantum many-body
systems described by translation-invariant Hamiltonians, sums of finite-range
interaction terms~\cite{lieb1972}. Specifically, they showed that
any \textit{causal} response function 
\begin{equation}
\chi_{AB}(\br,t)=-i\,\langle\Psi|\,\big[\mathcal{A}(\br,t),\mathcal{B}(\bnot,0)\big]\,|\Psi\rangle,\label{chi-AB}
\end{equation}
 with $t>0$ and arbitrary $|\Psi\rangle$, decays exponentially for
$|\br|>vt$ provided $\mathcal{A}(\br,t)$ and $\mathcal{B}(\br,t)$
are local operators in the Heisenberg form, i.e., such that $\big[\mathcal{A}(\br,t),\mathcal{B}(\bnot,t)\big]$
is non zero only for $\br=\bnot$. The velocity $v$ is finite and
can be upper estimated by a properly defined operator norm of each
local interaction term~\cite{lieb1972,nachtergaele2006}. The velocity
$v$ does not depend on the wave function $|\Psi\rangle$, but only
on the spectrum of the Hamiltonian. It is remarkable that, even though
$|\Psi\rangle$ may be highly entangled and possess long-range correlations,
any local perturbation needs a finite time to propagate up to a given
distance. Such a locality principle constitutes a rather fundamental
aspect in the dynamics of interacting many-body quantum systems, which
is attracting considerable attention in recent years, mainly sparked
by the impressive progress in ultracold-atom experiments. These experiments
allow for a direct access to the non-equilibrium dynamics of relatively
simple and quasi-isolated systems, making it possible to address issues
that until recently were considered merely academic~\cite{polkovnikov2011,lewenstein2007,bloch2008}.

A related question arises when one considers instead equal-time correlations
of the form 
\begin{equation}
N_{AB}(\br,t)=\langle\Psi|\,\mathcal{A}(\br,t)\mathcal{B}(\bnot,t)-\mathcal{A}(\br,0)\mathcal{B}(\bnot,0)\,|\Psi\rangle.\label{N-AB}
\end{equation}
 Although in $N_{AB}(\br,t)$ the \textsl{measurement} is instantaneous,
unlike in $\chi_{AB}(\br,t)$, several arguments suggest that an horizon
effect emerges even for $N_{AB}(\br,t)$, with a light-cone velocity
twice as large as the Lieb-Robinson bound~\cite{bravyi2006,calabrese2006}.
Early evidence of a light-cone effect in the dynamics induced by interaction
quenches in one-dimensional Bose-Hubbard models was found in Ref.~\cite{lauchli2008}
using time-dependent density-matrix renormalization group (tDMRG)~\cite{white2004,daley2004}
and confirmed experimentally in Ref.~\cite{cheneau2012}. However,
these first results raise intriguing questions that are worth investigating.
On the one hand, outside the Mott insulator phase, the bosons form
a superfluid with power-law correlations. The infinite correlation
length, alike a system right at criticality, would suggest that the
light-cone velocity is just once or twice the sound velocity (i.e.,
the velocity of the critical modes) for the correlation functions~(\ref{chi-AB})
and (\ref{N-AB}) respectively, as predicted by conformal-field theory
(CFT)~\cite{calabrese2006}. The t-DMRG analysis appears to call
into question the CFT prediction~\cite{lauchli2008}. However, accurate
determination of the propagation velocity and comparison to the characteristic
velocities of the system remain open questions. On the other hand,
the spreading of correlations in dimensions higher than one constitutes
an almost unexplored land, where t-DMRG approaches do not apply. This
question is particularly relevant in view of the possibility of extending
the experimental results~\cite{cheneau2012,Langen2013} in higher
dimensions.

In this paper we study these questions using the recently introduced
time-dependent Variational Monte Carlo (t-VMC) approach~\cite{carleo2012},
which allows us to address asymptotically long propagation times and
dimensions higher than one. Specifically we study the spreading of
density-density correlations after a quench in the interaction strength
of the Bose-Hubbard model in one (1D) and two (2D) dimensions. For
both cases in the superfluid regime, we find a supersonic light-cone
effect. More precisely, we find that the light-cone velocity differs
from both twice the sound velocity and twice the maximum excitation
velocity, except when approaching the Mott transition, where these
velocities are equal. Moreover, we show that in 2D the correlation
spreading is highly anisotropic and present nontrivial interference
effects. The anisotropy of the correlation front is however simply
explained in terms of the lattice coordination within the Manhattan
metrics.

\textit{System and method.---} We consider non-relativistic lattice
bosons described by the Bose-Hubbard Hamiltonian 
\begin{equation}
\mathcal{H}(U)=-\sum_{\langle\mathbf{R},\mathbf{R}^{\mathbf{\prime}}\rangle}\left(b_{\mathbf{R}}^{\dagger}b_{\mathbf{R}^{\prime}}^{\dagga}+\text{h.c.}\right)+\frac{U}{2}\sum_{\mathbf{R}}\, n_{\mathbf{R}}(n_{\mathbf{R}}-1),\label{eq:bosehubbard}
\end{equation}
 where $\mathbf{R}$ denotes a lattice site, $\langle\mathbf{R},\mathbf{R}^{\mathbf{\prime}}\rangle$
a pair of nearest-neighbor sites, $b_{\mathbf{R}}^{\dagger}$ ($b_{\mathbf{R}}^{\dagga}$)
the creation (annihilation) operator of a boson on site $\mathbf{R}$,
$n_{\mathbf{R}}=b_{\mathbf{R}}^{\dagger}b_{\mathbf{R}}^{\dagga}$
the boson density on site $\mathbf{R}$, and $U$ the two-body interaction
strength. In the following, the lattice will be either a 1D chain
or a 2D square lattice, with periodic boundary conditions and average
density $\langle n_{\bR}\rangle=1$. The system is first prepared
in the ground state of $\mathcal{H}(U_{\textrm{i}})$. At time $t=0$,
it is then driven out of equilibrium upon realizing a sudden quantum
quench in the interaction strength, from $U_{\textrm{i}}$ to $U_{\textrm{f}}$.
We study the dynamics of the density-density correlation function
\begin{equation}
N(\bR,t)=\langle n_{\bR}(t)n_{\bnot}(t)\rangle-\langle n_{\bR}(0)n_{\bnot}(0)\rangle,
\end{equation}
 where the average is over the ground state of $\mathcal{H}(U_{\textrm{i}})$
and the density operators are evolved in time with $\mathcal{H}(U_{\textrm{f}})$
i.e., Eq.~(\ref{N-AB}) where both $\mathcal{A}$ and $\mathcal{B}$
are the density operators.

Our analysis makes use of the t-VMC approach~\cite{carleo2012} that
we briefly outline here. The starting point is to define a class of
time-dependent variational many-body wave functions, which we take
of the Jastrow type 
\begin{equation}
\Psi(\mathbf{x},t)\equiv\langle\mathbf{x}|\Psi(t)\rangle=\exp\left[\sum_{r}\alpha_{r}(t)\mathcal{O}_{r}(\mathbf{x})\right]\Phi_{0}(\mathbf{x}),\label{eq:psivar}
\end{equation}
 where $\mathbf{x}$ spans a configuration basis, $\Phi_{0}(\mathbf{x})$
is a bosonic time-independent state, and $\alpha_{r}(t)$ are complex
variational parameters coupled to a set of operators $\mathcal{O}_{r}$
that are diagonal in the $\mathbf{x}$-basis, i.e., $\langle\mathbf{x}|\mathcal{O}_{r}|\mathbf{x^{\prime}}\rangle=\delta_{\mathbf{x},\mathbf{x^{\prime}}}\mathcal{O}_{r}(\mathbf{x})$.
The explicit form of these operators and their total number define
the variational subspace. Here we use the Fock basis, $\mathbf{x}=\{n_{i}\}$,
and the complete set of density-density correlations, $\mathcal{O}_{\br}=\sum_{\bR}n_{\bR}n_{\bR+\br}$,
where $\br$ spans all independent distances on the lattice. The initial
state is chosen to be the variational Jastrow ground state of $\mathcal{H}(U_{\textrm{i}})$
with $|\Phi_{0}\rangle$ the noninteracting-boson ground state of
$\mathcal{H}(0)$. This choice provides an excellent approximation
of the exact ground state of $\mathcal{H}(U_{\textrm{i}})$~\cite{capello2007,capello2008}.
For instance, the superfluid-insulator transition is obtained for
$U_{\textrm{c}}^{\text{var}}\simeq5$ and $U_{\textrm{c}}^{\text{var}}\simeq21$
in 1D and 2D respectively, in fair agreement with exact results~\cite{khuner1998,capogrosso2008}.

The variational dynamics of the system is fully contained in the trajectories
of the variational parameters $\alpha_{r}(t)$. The latter are obtained
by minimizing the Hilbert-space distance between the infinitesimal
\textit{exact} dynamics and the time derivative of the variational
state~(\ref{eq:psivar}) at each time step. This process is equivalent
to project the exact time-evolved wave function onto the variational
subspace. It yields a closed set of coupled equations of motion: 
\begin{equation}
i\sum_{r^{\prime}}S_{r,r^{\prime}}(t)\overset{\bm{.}}{\alpha}_{r^{\prime}}(t)=\langle\mathcal{O}_{r}\mathcal{H}\rangle_{t}-\langle\mathcal{O}_{r}\rangle_{t}\langle\mathcal{H}\rangle_{t},\label{eq:syseq}
\end{equation}
 where $S_{r,r^{\prime}}(t)=\langle\mathcal{O}_{r}\mathcal{O}_{r^{\prime}}\rangle_{t}-\langle\mathcal{O}_{r}\rangle_{t}\langle\mathcal{O}_{r^{\prime}}\rangle_{t}$
and the quantum averages are taken over the time-dependent variational
state~(\ref{eq:psivar}). At each time, the quantum averages appearing
in Eq.~(\ref{eq:syseq}) are computed by variational Monte Carlo
simulations and the linear system of equations~\eqref{eq:syseq}
is solved for $\overset{\bm{.}}{\alpha}_{r}(t)$. The trajectories
$\alpha_{r}(t)$ are then found by time-integrating the functions
$\overset{\bm{.}}{\alpha}_{r}(t)$.

We emphasize that our variational scheme is symplectic and exactly
conserves both the total energy and the square modulus of the wave
function. In the numerical calculations, we use a sufficiently small
time-step, $\delta t=0.01$, and a fourth-order Runge-Kutta integration
scheme, which conserves the energy with a very small systematic error
of the order of one part in a thousand, for times up to $t=100$.
The t-VMC is therefore intrinsically stable, amenable to simulating
time scales that exceed by about two orders of magnitudes those achievable
by t-DMRG in 1D, and applies as well in higher dimensions.

%%%%%%%%%%%%%%%%%%%%%%%%%%%%%%%%%%%%%%%%%%%%%%%%%%%%%%%%%%%%%%%%%%%%%%%%%%
\begin{figure}[t]

\noindent \begin{centering}
\includegraphics[bb=0bp 0bp 415bp 434bp,clip,width=1\columnwidth]{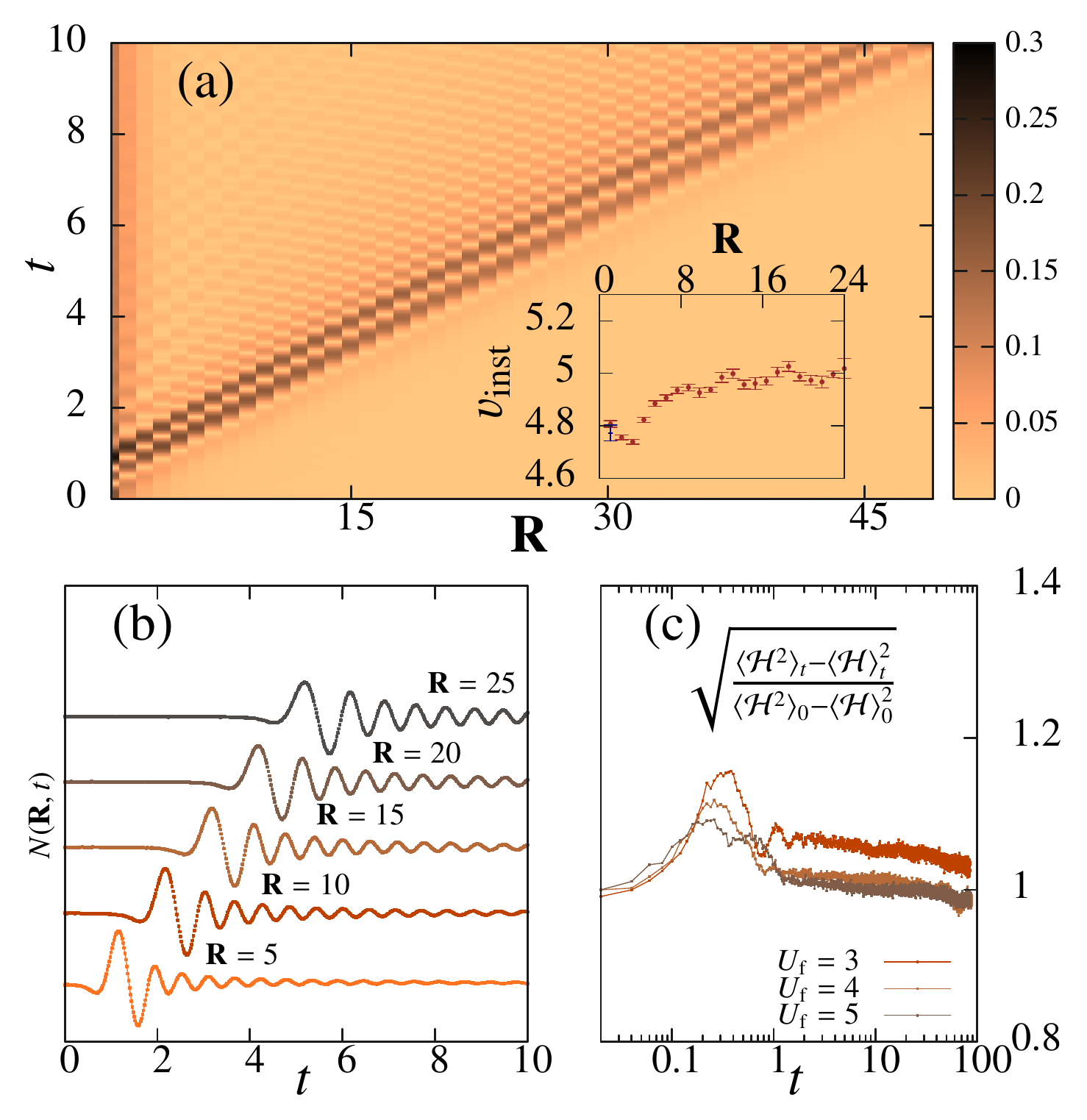} 
\par\end{centering}

\caption{\label{fig:SpCorr} (Color on-line) Spreading of correlations in a
1D chain. (a) Density-density correlations $N(R,t)$ versus separation
and time for a quench in the interaction strength from $U_{\textrm{i}}=2$
to $U_{\textrm{f}}=4$. The inset shows the instantaneous velocity
as obtained from t-VMC (red points) and exact diagonalization (for
a 12-site lattice; blue point). (b) Time dependence of $N(R,t)$ for
various values of $R$. For clarity, the curves are vertically shifted
by a value proportional to $R$, and the linear light-cone wave-front
clearly appears. (c) Relative energy fluctuations versus time for
various values of $U_{\textrm{f}}$. The t-VMC calculations are performed
for 200 (a and b) or 500 (c) sites.}
\end{figure}

%%%%%%%%%%%%%%%%%%%%%%%%%%%%%%%%%%%%%%%%%%%%%%%%%%%%%%%%%%%%%%%%%%%%%%%%%%

\textit{Results.---} Let us first discuss our results for the 1D chain.
Figure~\ref{fig:SpCorr}(a) shows the density-density correlation
$N(R,t)$ as a function of separation and time for a quantum quench
from $U_{\textrm{i}}=2$ to $U_{\textrm{f}}=4$. Figure~\ref{fig:SpCorr}(b)
shows vertical cuts of the latter, plotted with a vertical shift proportional
to $R$ for clarity. A light-cone effect is clearly visible: $N(R,t)$
is unaffected at short times, then develops a maximum at a finite
time $t^{\star}(R)$, and finally undergoes damped oscillations. Similar
results are found for all quenches discussed below. For large enough
separation, the activation time $t^{\star}(R)$ depends linearly on
the separation, $t^{\star}(R)\equiv v_{\textrm{lc}}\times R$, which
defines the light-cone velocity $v_{\text{lc}}$. More precisely,
the instantaneous correlation-spreading velocity, $v_{\text{inst}}(R)\equiv\frac{2}{t^{\star}(R+1)-t^{\star}(R-1)}$,
is shown as a function of $R$ in the inset of Fig.~\ref{fig:SpCorr}(a).
The ballistic regime, where $v_{\text{inst}}(R)$ approaches $v_{\textrm{lc}}$,
is achieved only for sufficiently long time ($t_{\text{ball}}\sim4$).
The t-VMC method allows us to simulate very long times in the asymptotic
ballistic regime ($t\sim100$), and extract accurate values of $v_{\text{lc}}$.

At variance with the total energy, higher moments of the Hamiltonian
are not strictly conserved by the t-VMC scheme, as illustrated in
Fig.~\ref{fig:SpCorr}(c). Nevertheless, despite a slight time-dependence
of the energy fluctuations at very short times, the long-time value
always coincides with the initial value, showing the accuracy of our
variational method. In order to further check it, we compared our
results (red points) to exact diagonalization (blue point) at time
$t\simeq0.5$ close to the maximal deviation of the energy fluctuations
(inset of Fig.~\ref{fig:SpCorr}(a)). We found very good agreement,
hence confirming the accuracy of t-VMC %
\footnote{In addition, we have compared the t-VMC velocities with the ones obtained
by t-DMRG in Ref.\cite{lauchli2008}. The two results are in quantitative
agreement, within the error bars of the t-dmrg calculation.%
}.

The very existence of a finite propagation velocity and its microscopic
origin can be justified as follows. Assume $|n\rangle$ and $|m\rangle$
are two eigenstates of $\mathcal{H}(U_{\textrm{f}})$ with eigenvalues
$E_{l}$ and $E_{m}$, and total momentum $\mathbf{P}+\bq$ and $\mathbf{P}$,
respectively, such that $\langle l|\,\mathcal{A}_{\bq}\,|m\rangle$,
with $\mathcal{A}_{\bq}=\sum_{\br}\mathcal{A}(\br)\text{e}^{i\bq\cdot\br}$,
is finite. If $\mathcal{A}(\br)$ is a bounded local operator, then
$\omega_{lm}(\bP,\bq)=E_{l}-E_{m}$ is not an extensive quantity,
though $E_{l}$ and $E_{m}$ are both extensive. For large $\br$,
i.e.\ small $\bq$, such excitation can propagate coherently only
if $|\br|\simeq t\,|\boldsymbol{\partial}_{\bq}\,\omega_{lm}(\bP,\bnot)|$.
This defines a maximum propagation velocity $v_{\text{m}}\equiv\text{Max}|\boldsymbol{\partial}_{\bq}\,\omega_{lm}(\bP,\bnot)|$
to be identified with the Lieb-Robinson bound. In the case of Eq.~\eqref{N-AB},
if $|\Psi\rangle$ is an eigenstate of defined total momentum, then
two counter-propagating excitations are involved due to momentum conservation,
and the bound velocity is $2v_{\text{m}}$. For small quenches towards
a gapless phase, one may expect that only low-energy phonon excitations
are involved, and that the light-cone velocity is twice the sound
velocity, $2v_{\text{s}}$.

%%%%%%%%%%%%%%%%%%%%%%%%%%%%%%%%%%%%%%%%%%%%%%%%%%%%%%%%%%%%%%%%%%%%%%%%%%
\begin{figure}
\noindent \begin{centering}
\includegraphics[bb=7bp 5bp 349bp 209bp,clip,width=1\columnwidth]{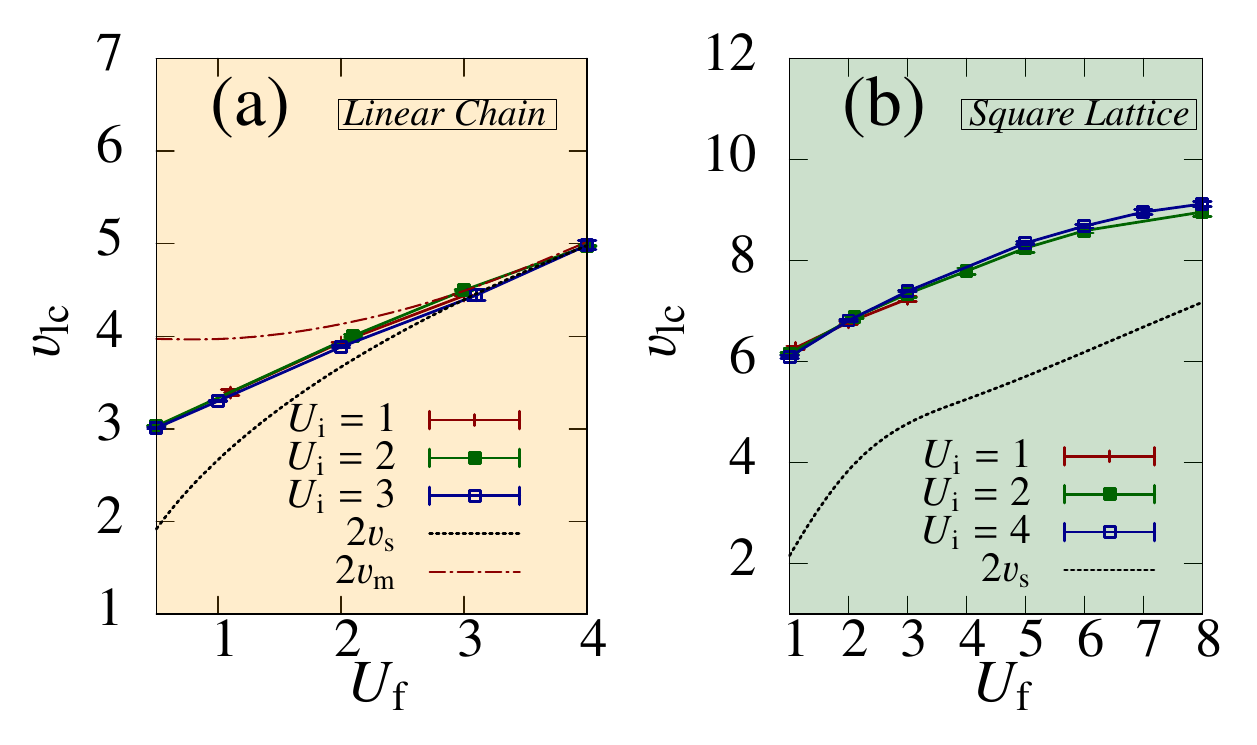} 
\par\end{centering}

\caption{\label{fig:LR-vel} (Color on-line) Light-cone velocity $v_{\text{lc}}$
versus the final interaction strength $U_{\textrm{f}}$, for various
values of the initial interaction strength $U_{\textrm{i}}$. (a)
1D chain. (b) 2D square lattice. Also shown are twice the sound velocity,
$2v_{\text{s}}$, and twice the maximum excitation velocity, $2v_{\text{m}}$,
for $\mathcal{H}(U_{\textrm{f}})$. }
\end{figure}

%%%%%%%%%%%%%%%%%%%%%%%%%%%%%%%%%%%%%%%%%%%%%%%%%%%%%%%%%%%%%%%%%%%%%%%%%%

The value of $v_{\text{lc}}$ is plotted in Fig.~\ref{fig:LR-vel}(a)
as a function of the final interaction strength $U_{\textrm{f}}$
for various values of initial interaction strength $U_{\textrm{i}}$.
We find that $v_{\text{lc}}$ increases with $U_{\textrm{f}}$, which
is readily understood by the fact that the rigidity of the final lattice
increases with $U_{\textrm{f}}$. It is remarkable however that $v_{\text{lc}}$
does not depend on $U_{\textrm{i}}$. In Fig.~\ref{fig:LR-vel},
the t-VMC value for $v_{\text{lc}}$ is compared to the characteristic
velocities of the density excitations, i.e. $2v_{\text{m}}$ and $2v_{\text{s}}$.
The latter ones are computed as $v_{\text{m}}=\max\{\partial E(q)/\partial q\}$
and $v_{\text{s}}=\lim_{q\rightarrow0}\partial E(q)/\partial q$,
where $E(q)$ is the energy of the density modes $\left|\psi(q)\right\rangle =\rho(q)\left|\psi_{0}\right\rangle $,
with $\rho(q)$ the Fourier transform of the density operator~\cite{notasound,capello2007}.
We generically find that the light-cone velocity significantly differs
from twice both these velocities. On the one hand, the maximum velocity
allowed by the propagation of excitations is not achieved, in contrast
to quenches from the Mott phase~\cite{cheneau2012,barmettler2012}.
On the other hand, a supersonic regime is achieved in all the superfluid
region of the out-of-equilibrium phase diagram, even for very small
quenches. For instance, in the case of a quench from $U_{\textrm{i}}=1$
to $U_{\textrm{f}}=1.1$, we find $v_{\text{lc}}=3.39(3)$ and $2v_{\text{s}}=2.78$
{[}see Fig.~\ref{fig:LR-vel}(a){]}. Therefore, high-energy excitations
beyond the sound-wave regime are always generated by the quench dynamics.
It corresponds to short-distance effects that are always significant
but not accounted for in CFT~\cite{calabrese2006}. A form of universality
is recovered only in the neighborhood of the Mott transition. When
the final interaction strength approaches the critical value, $U_{\textrm{c}}^{\text{var}}\simeq5$
at the variational level, the excitation modes exhibit a maximal velocity
at zero momentum and all the characteristic velocities, $v_{\text{lc}}$,
$2v_{\text{m}}$, and $2v_{\text{s}}$ coincide. It suggests that
the results of CFT are correct only when the quantum quench is performed
right at a critical point and not in the whole quasi-long-range ordered
phase with infinite correlation length, i.e., for $U_{\textrm{f}}<U_{\textrm{c}}^{\text{var}}$.

\begin{figure}
\noindent \begin{centering}
\includegraphics[bb=0bp 0bp 343bp 117bp,clip,width=1\columnwidth]{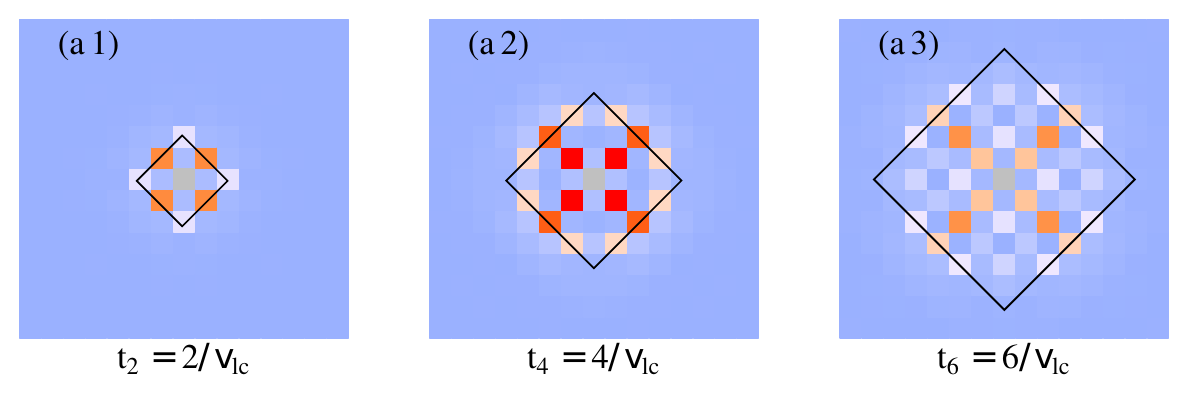} 
\par\end{centering}

\noindent \begin{centering}
\includegraphics[width=1\columnwidth]{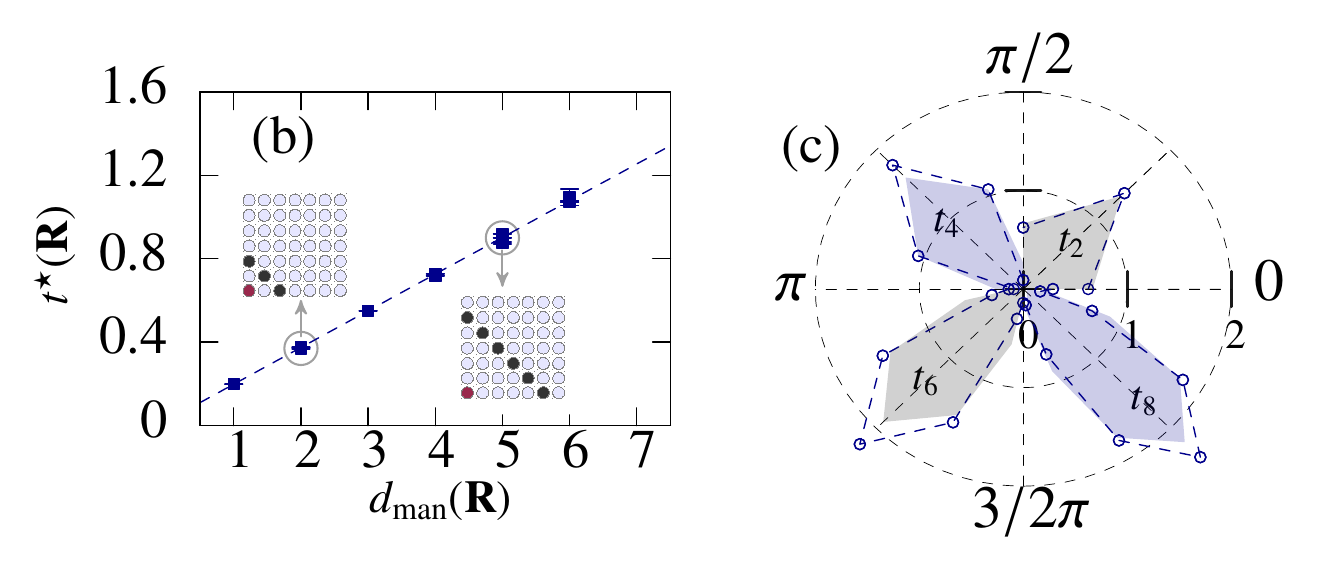} 
\par\end{centering}

\caption{\label{fig:SpCorr2d} (Color on-line) Spreading of correlations in
a $20\times20$-site square lattice for a quench from $U_{\textrm{i}}=2$
to $U_{\textrm{f}}=4$. (a)~Density-density correlations $N(\bR,t)$
at fixed times $t_{n}=n/v_{\text{lc}}$. The $45^{\circ}$-tilted
squares denote the points on the correlation front. (b)~Activation
time $t^{\star}(\bR)$ versus Manhattan distance $d_{\textrm{man}}(\bR)$
for various points $\bR$. The insets show the ensemble of points
$\bR$ with equal $d_{\textrm{man}}(\bR)$ at the corresponding Manhattan
distance. The dashed line is a linear fit to the data. (c)~Intensity
of the correlation signal (dots) and number of paths (shaded areas)
versus the azimuthal angle of the points on correlation fronts. Each
quadrant corresponds to a polar plot at the $4$ different times indicated
on the Figure.}
\end{figure}

We now turn to the 2D square lattice. The spreading of correlations
in dimension higher than one constitutes an almost unexplored land
where only mean-field methods have been applied so far~\cite{menotti2007,sotiriadis2010,sciolla2010,schiro2010,Carusotto,Natu}.
The latter are reliable only in the unphysical limits of large lattice
connectivity or large internal {}``flavor'' degeneracy. In contrast,
t-VMC takes into account relevant dynamical correlations and can be
applied to the physical Bose-Hubbard Hamiltonian in any dimension.
Figure~\ref{fig:SpCorr2d}(a) shows the correlation function $N(\bR,t)$
at equally separated times, for a quench from $U_{\textrm{i}}=2$
to $U_{\textrm{f}}=4$ in the 2D square lattice. It shows a clear
spreading of correlations. The correlation front is a square with
principal axes along the diagonals of the lattice. In order to understand
this, notice that nearest-neighbor hopping in the square lattice induces
a natural metrics that is of the Manhattan type~\cite{manhattan},
rather than Euclidean. Points at equal Manhattan distance $d_{\text{man}}(\bR)\equiv|R_{x}|+|R_{y}|$
are thus located on $45^{\circ}$-tilted squares. Figure~\ref{fig:SpCorr2d}(b)
shows the activation time $t^{\star}(\bR)$, defined as the time when
the first maximum of $N(\bR,t)$ appears, versus the Manhattan distance
for various lattice sites. The data for various $\bR$ but same $d_{\text{man}}(\bR)$
collapse, which confirms that the Manhattan distance is the relevant
metrics. Moreover, within the Manhattan metrics, a clear ballistic
behavior is observed, which allows us to define the light-cone velocity
$v_{\text{lc}}\equiv d_{\text{man}}(\bR)/t^{\star}(\bR)$. In Fig.~\ref{fig:LR-vel}(b),
we show the extracted values of $v_{\text{lc}}$ as a function of
$U_{\textrm{f}}$ for various values of $U_{\textrm{i}}$, together
with twice the sound velocity for the 2D square lattice. As for the
1D chain, a strong discrepancy between these two velocities is found
also in 2D. This outcome indicates that high-energy excitations dominate
the dynamical evolution even for small quenches, although the initial
state is genuinely off-diagonal long-range ordered. It contrasts with
low-enegy descriptions that take into account only sound modes.

%%%%%%%%%%%%%%%%%%%%%%%%%%%%%%%%%%%%%%%%%%%%%%%%%%%%%%%%%%%%%%%%%%%%%%%%%%

%%%%%%%%%%%%%%%%%%%%%%%%%%%%%%%%%%%%%%%%%%%%%%%%%%%%%%%%%%%%%%%%%%%%%%%%%%

As it can be seen on Fig.~\ref{fig:SpCorr2d}(a), the correlation
signal shows complicated, anisotropic patterns, as a result of nontrivial
interference effects. For instance at variance with the 1D case, the
time-dependence of $N(\bR,t)$ can show several secondary maxima with
a stronger amplitude than the wavefront. The anisotropy can however
be understood on the wavefront where the interference effects are
weak. Indeed, two points $(0,0)$ and $\bR=(R_{x},R_{y})$ are generically
connected by a number $N_{\text{man}}(\bR,d)$ of paths of total length
$d$, which do not depend only on $d_{\text{man}}(\bR)$. On the wavefront,
$d=d_{\text{man}}(\bR)$ and $N_{\text{man}}[\bR,d]=(|R_{x}|+|R_{y}|)!/|R_{x}|!|R_{y}|!$,
which grows from $1$ on the angles to $d!/[(d/2)!]^{2}$ on the center
of the sides. This explains that the maxima are located on the main
axis of the correlation square. More precisely, Fig.~\ref{fig:SpCorr2d}(c)
shows both the intensity of the correlation signal (points) and the
number of connecting paths $N_{\text{man}}[\bR,d]$ (shaded areas),
for various times and various points on the wavefront. The quantitative
agreement between the two confirms that the main source of anisotropy
on the correlation front is geometrical.

\textit{Conclusions.---} We have studied the spreading of density-density
correlations after a quantum quench in 1D and 2D Bose-Hubbard models,
using the recently developed t-VMC approach. Our results show a light-cone
ballistic expansion of correlations in both cases, and provide accurate
values of the light-cone velocity. Our main result is that the light-cone
velocity significantly differs from both twice the sound velocity
and twice the maximum excitation velocity, except when approaching
the Mott transition. Moreover, in 2D, the correlation signal is highly
anisotropic and the correlation front is a square, which is due to
the Manhattan metrics imposed by the nearest-neighbor lattice coordination.
Our results provide new insight on the spreading of correlations in
interacting quantum systems. They also offer an important benchmark
for future experiments with ultracold atomic gases in optical lattices,
especially in dimension higher than one.
\begin{acknowledgments}
We acknowledge discussions with I. Bouchoule, I. Carusotto, M. Cheneau,
and M. Schiro. This research was supported by the European Research
Council (FP7/2007-2013 Grant Agreement No.~256294), Marie Curie IEF
(FP7/2007-2013 - Grant Agreement No.~327143) and PRIN 2010-11. Use
of the computing facility cluster GMPCS of the LUMAT federation (FR
LUMAT 2764) is acknowledged. 
\end{acknowledgments}
\textbf{ }


\begin{thebibliography}{References}
\bibitem{lieb1972} E.H. Lieb and D.W. Robinson, Commun. Math. Phys.
\textbf{28}, 251 (1972).

\bibitem{nachtergaele2006} B. Nachtergaele and R. Sims, Commun. Math.
Phys. \textbf{265}, 119 (2006).

\bibitem{polkovnikov2011} A. Polkovnikov, K. Sengupta, A. Silva,
and M. Vengalattore, \rmp \textbf{83}, 863 (2011).

\bibitem{lewenstein2007} M. Lewenstein, A. Sanpera, V. Ahufinger,
B. Damski, A. Sen De, and U. Sen, Adv. Phys. \textbf{56} 243 (2007).

\bibitem{bloch2008} I. Bloch, J. Dalibard, and W. Zwerger, \rmp
\textbf{80}, 885 (2008).

\bibitem{bravyi2006} S. Bravyi, M.B. Hastings, and F. Verstraete,
\prl \textbf{97}, 050401 (2006).

\bibitem{calabrese2006} P. Calabrese and J. Cardy, \prl \textbf{96},
136801 (2006); J. Stat. Mech. P06008 (2007).

\bibitem{lauchli2008} A.M. Lauchli and C. Kollath, J. Stat. Mech.
P05018 (2008).

\bibitem{white2004} S.R. White and A.E. Feiguin, \prl \textbf{93},
076401 (2004).

\bibitem{daley2004} A.J. Daley, C. Kollath, U. Schollwoeck, and G.
Vidal, J. Stat. Mech. P04005 (2004).

\bibitem{cheneau2012} M. Cheneau, P. Barmettler, D. Poletti, M. Endres,
P. Schauss, T. Fukuhara, C. Gross, I. Bloch, C. Kollath, and S. Kuhr,
Nature \textbf{481}, 484 (2012).

\bibitem{Langen2013} T. Langen, R. Geiger, M. Kuhnert, B. Rauer,
and J. Schmiedmayer, Nat. Phys. \textbf{9}, 640 (2013).

\bibitem{carleo2012} G. Carleo, F. Becca, M. Schiro, and M. Fabrizio,
Sci. Rep. \textbf{2}, 243 (2012).

\bibitem{capello2007} M. Capello, F. Becca, M. Fabrizio, and S. Sorella,
\prl \textbf{99}, 056402 (2007).

\bibitem{capello2008} M. Capello, F. Becca, M. Fabrizio, and S. Sorella,
\prb \textbf{77}, 144517 (2008).

\bibitem{khuner1998} T.D. Kuhner and H. Monien, \prb \textbf{58},
14741 (1998).

\bibitem{capogrosso2008} B. Capogrosso-Sansone, S.G. Soyler, N. Prokof'ev,
and B. Svistunov, \pra \textbf{77}, 015602 (2008).

\bibitem{notasound} The sound velocity may be computed by considering
the size corrections of the ground-state energy.~\cite{capello2008}
In one dimension, the Lieb-Liniger result $v_{LL}=2\sqrt{\gamma-\gamma^{3/2}/2\pi}$,
with $\gamma=U/2J$, ~\cite{lieb1963} gives an excellent approximation
to the sound velocity.

\bibitem{lieb1963} E.H. Lieb, Phys. Rev. \textbf{130}, 1616 (1963).

\bibitem{barmettler2012} P. Barmettler, D. Poletti, M. Cheneau, and
C. Kollath, \pra 85, 053625 (2012).

\bibitem{menotti2007} C. Menotti, C. Trefzger, and M. Lewenstein,
\prl \textbf{98}, 235301 (2007).

\bibitem{sotiriadis2010} S. Sotiriadis and J. Cardy, Phys. Rev. B
\textbf{81}, 134305 (2010).

\bibitem{sciolla2010} B. Sciolla and G. Biroli, \prl \textbf{105},
220401 (2010).

\bibitem{schiro2010} M. Schiro and M. Fabrizio, \prl \textbf{105},
076401 (2010).

\bibitem{Carusotto} I. Carusotto, R. Balbinot, A. Fabri and A. Recati,
Eur. Phys. J. D\textbf{ 5}6, 391 (2010).

\bibitem{Natu} S.S. Natu, and E.J. Mueller, \pra \textbf{87}, 063616
(2013).

\bibitem{manhattan} E. F. Krause, \emph{Taxicab Geometry, }(Courier
Dover Publications, 1986).\end{thebibliography}
\end{document}